\documentclass[aps,prd,eqsecnum,amsmath,nofootinbib,preprintnumbers]{revtex4}%
\usepackage{amsfonts,amssymb}
\usepackage{graphicx}
\usepackage{xcolor}
\usepackage{float}
\definecolor{wred}{rgb}{0,0.618,0.0}
\definecolor{wblue}{rgb}{.0,0.0,0.618}
\definecolor{wgreen}{rgb}{.0,0.618,0.0}
\usepackage{tikz}
\usepackage{multirow}
\usetikzlibrary{decorations.pathmorphing,calc}
\usepackage{bm}
\usepackage[
colorlinks=true,
filecolor=black,
linkcolor=wblue,
citecolor=wgreen,
pdfstartview=FitH,
pdfpagemode=None,
bookmarksopen=true,
urlcolor=blue%{blue!80!black!80}
]{hyperref}
\begin{document}
	\preprint{\hfill {\small {1}}}
\title{Matter field and black hole horizon geometry}
\author{Yuxuan Peng}
\email{yxpeng@alumni.itp.ac.cn}
\affiliation{Department of Physics, School of Science, East China University of Technology, Nanchang, Jiangxi 330013, P.R. China}
\affiliation{}

\begin{abstract}

This paper investigates the influence of matter fields on the geometry of black hole horizons within higher-order gravity theories. Focusing on five-dimensional Einstein-Gauss-Bonnet gravity at a critical coupling constant ($\alpha = -3/(4\Lambda)$), we demonstrate that while vacuum solutions permit horizons with arbitrary geometry, the introduction of a scalar field imposes constraints. Specifically, the scalar hair restricts the horizon to manifolds of constant scalar curvature, extending beyond the Thurston geometries (Sol, Nil, $SL_2R$) previously identified. We prove a uniqueness theorem for the scalar-coupled solutions, showing that the metric and scalar field must adopt specific forms, with the horizon geometry solely required to satisfy the constant curvature condition. Furthermore, analogous results are established in generic $F(R)$ gravity, where arbitrary horizons with constant scalar curvature emerge at critical couplings, exemplified by $F(R) = R + \lambda R^2 - 2\Lambda$ with $\lambda = -1/(8\Lambda)$. These findings highlight a probably universal feature: critical couplings in higher-order gravity enable unconstrained horizon geometries in vacuum, while matter fields introduce geometric restrictions. This work deepens understanding of the interplay between matter, higher-curvature corrections, and black hole horizon geometry.

\end{abstract}

\maketitle

\section{Introduction}

Black holes are big secrets in the universe, and their horizons are extremely mysterious: hardly visible, deeply hidden in strong gravity field and swallowing nearly everything.
In theoretical models, static black hole horizons often appear in spherical shape (the famous Schwarzschild and Reissner-N\"ordstrom and many other examples) or other maximally symmetric shapes (asymptotically Anti-de Sitter (AdS) black holes with flat or hyperbolic horizons\cite{Birmingham:1998nr}).
These black hole solutions can be included in the following generic expression of ``warped product'' in $D=(m+n)$ dimensions
\begin{eqnarray}\label{decomp}
    \mathrm{d}s^2 = g_{ab}(u)\mathrm{d}u^a\mathrm{d}u^b +\gamma^2(u)\hat{g}_{ij}(x)\mathrm{d}x^i\mathrm{d}x^j\,.
\end{eqnarray}
The $g_{ab}$ is the metric of the $m$-dimensional part and $u$ can be $t$ and $r$ for some simple solutions, while $\hat{g}_{ij}(x)$ is the metric of the $n$-dimensional base manifold (horizon cross section while at $r=r_h$) with coordinates $x^i$.
The latin indices $a,b,c,...$ and $i,j,k,...$ denotes the spacetime indices of the $m$-dimensional submanifold and the $n$-dimensional base manifold respectively.
The geometric quantities on the base manifold all appear with a $\hat{}$ on them, and the geometric quantities of the $m$-dimensional submanifold with $g_{ab}$ are denoted by a superscript $\bar{}$. The decomposition of the curvature quantities is shown in detail in the appendix.

For static case the metric is further simplified to the expression below,
\begin{eqnarray}\label{staticdecomp}
    \mathrm{d}s^2 = -V(r)\mathrm{d}t^2 + \frac{1}{V(r)}\mathrm{d}r^2 +r^2\hat{g}_{ij}\mathrm{d}x^i\mathrm{d}x^j\,,
\end{eqnarray}
and $\hat{g}_{ij}$ can be spherical, flat, hyperbolic or other.
People already know\cite{GibWil87} that for Einstein gravity with a cosmological constant, $\hat{g}_{ij}$ can be a metric of an n-dimensional Einstein manifold --- $\hat{R}_{ij} \propto \hat{g}_{ij}$.
An interesting problem is to explore other possibilities --- which we call ``nontrivial black hole horizons'' --- other than the aforementioned maximally symmetric or Einstein manifolds.
For 4-dimensional black holes, in the presence of specific matter fields, the horizon geometry is indeed enriched, shown by Yang's work \cite{Yang:2023nnk} in 2023.
In that paper 2 novel topological black hole exact solutions with unusual shapes of horizons in the simplest holographic axions model, the four-dimensional Einstein-Maxwell-axions theory, are constructed.
In 5 dimensions, the horizon is 3-dimensional, possessing richer structures than 2-dimensional ones.
For example, according to their symmetry, homogeneous 3-dimensional manifolds can be classified  into 9 Bianchi types\cite{Ryan:1975jw}, or into 8 Thurston geometries \cite{Th1,Sc1}.
Thurston's classification includes the Euclidean $E^3$, spherical$S^3$, hyperbolic $H^3$, $S^1\times S^2$, $S^1\times H^2$ and the so-called ``Sol'', ``Nil'' and ``$SL_2 R$'' geometries.
Cadeau and Woolgar\cite{Cadeau:2000tj} found out vacuum static black hole solutions with Sol and Nil horizons in Einstein gravity with a negative cosmological horizon, however not in the warped product form as (\ref{staticdecomp}).
Afterwards the Thurston horizons were studied in many different settings\cite{Hervik:2003vx,Arias:2017yqj,Arias:2018mqn,Faedo:2019rgo,Figueroa:2021apr,Peng:2021xwh,Naderi:2021skd,Faedo:2022hle}, enriching our knowledge of black hole horizons a lot.
% 加入引用: Thurston horizon 和 bianchi horizon 一系列进展(包括我本人的)

Higher curvature corrections may help constructing nontrivial horizons in warped forms.
Dotti et al.\cite{Dotti:2007az,Dotti:2010bw,Oliva:2012ff} studied the famous Einstein-Gauss-Bonnet gravity model with a critical coupling constant, the so-called ``Chern-Simons'' gravity, and found that there exist black hole horizons with arbitrary geometry.
Guajardo and Oliva\cite{Guajardo:2024hrl} added a scalar field into exactly those black hole solutions, and found that the horizon geometry is not totally arbitrary any more.
They proposed several cases for the horizon geometry, and they are just the Sol, Nil and $SL_2 R$ horizons mentioned above.
This result is very interesting and this paper explores this topic a bit further, in order to discuss the effect of matter fields on black hole horizon structure in the future.

In Sec.\ref{GBsection}, we explain in detail how the horizon geometry becomes arbitrary in the vacuum case, and how this geometry get restricted after the scalar matter is added. Actually the scalar field only requires the horizon
to have constant scalar curvature, not necessarily to be the Thurston geometries.
We also prove a uniqueness theorem of this model, stating that the metric and the scalar field can only appear in the form proposed by \cite{Guajardo:2024hrl}, except that the horizon geometry can be beyond Thurston geometries.
In Sec.\ref{sectionofFR}, we show a similar case in generic vacuum $F(R)$ theory where the black hole horizon is unconstrained, and give an explicit example of $F(R) = R + \lambda R^2 - 2\Lambda$ where the coupling constant $\lambda$ is at a critical value.
In the end we make the conclusion.

\section{Einstein-Gauss-Bonnet black hole with scalar hair}\label{GBsection}

The recent paper \cite{Guajardo:2024hrl} by Guajardo and Oliva introduces scalar hair to the 5-dimensional Einstein-Gauss-Bonnet gravity action to get static hairy black hols solutions with nontrivial horiozon geometry.
The action in that paper is
\begin{eqnarray}\label{GBactionMat}
   S &= &\int \mathrm{d}^{5} x \sqrt{-g}\left[ \frac{1}{2\kappa}( R - 2\Lambda + \alpha (R^2 -4 R_{\mu\nu}R^{\mu\nu} + R_{\mu\nu \rho \sigma}R^{\mu\nu \rho \sigma}))  - \dfrac{1}{2}\nabla_{\mu}\phi\nabla^{\mu}\phi - \dfrac{3}{32}\xi R\phi^2 - U(\phi) \right]
\end{eqnarray}
with $\Lambda<0$.
The system includes a scalar field $\phi$ and a nonminimal coupling $R\phi^2$.
The equations of motion include the gravitational field equations
\begin{eqnarray}
        &&R_{\mu\nu} - \frac{1}{2}R g_{\mu\nu} + \Lambda g_{\mu\nu}\nonumber\\
        &&- \alpha\left(4 R_{\mu\rho} R^\rho\,_\nu - 2R R_{\mu\nu}+ 4 R^{\rho \sigma}R_{\mu\rho\nu\sigma} - 2 R_\mu\,^{\rho\sigma\tau}R_{\nu\rho\sigma\tau} + \frac{1}{2} g_{\mu\nu} (R^2 -4 R_{\rho \sigma}R^{\rho \sigma} + R_{\rho\sigma\tau\pi}R^{\rho\sigma\tau\pi})\right)\nonumber\\
        && = \kappa\Big\{ \nabla_{\mu}\phi\nabla_{\nu}\phi - g_{\mu\nu}\left(\frac{1}{2}\nabla_{\sigma}\phi\nabla^{\sigma}\phi + U(\phi) \right) + \dfrac{3}{16} \xi \left(g_{\mu\nu}\square - \nabla_{\mu}\nabla_{\nu} + G_{\mu\nu} \right)\phi^2\Big\}\,\label{GBeomphi}
\end{eqnarray}
and the scalar field equations
\begin{eqnarray}
 \square \phi = \dfrac{3}{16}R\phi + U'(\phi)\,.\label{phieom} 
\end{eqnarray}
In the vacuum case there is a solution shown by the paper \cite{Dotti:2007az} at the special coupling$\alpha=-3/(4\Lambda)$:
\begin{eqnarray}\label{BHvacuum}
    &&\mathrm{d}s^2 = -V(r)\mathrm{d}t^2 + \frac{1}{V(r)}\mathrm{d}r^2 +r^2\hat{g}_{ij}\mathrm{d}x^i\mathrm{d}x^j\nonumber\\
        &&V(r)=\frac{r^2}{4\alpha} - \mu\,, \qquad  \text{$\mu$ is an arbitrary constant and $\hat{g}_{ij}$ an arbitrary function of $x^i$.}
    \end{eqnarray}
The arbitrary constant $\mu$ implies that this can be a black hole solution. \textbf{In the vacuum case the free base manifold metric $\hat{g}(x)_{ij}$ means that the 3-dimensional black hole horizon has arbitrary geometry.}
{At the point $\alpha=-3/(4\Lambda)$, the $2$ AdS vacuums of the Einstein-Gauss-Bonnet theory merge to one, and the theory is often referred to as the Chern-Simons theory \cite{Chamseddine:1989nu,Brihaye:2013vsa}.}

The solutions given by \cite{Guajardo:2024hrl} with an $r$-dependent $\phi(r)$ is:
\begin{eqnarray}\label{BHScalar}
&&\mathrm{d}s^2 = -V(r)\mathrm{d}t^2 + \frac{1}{V(r)}\mathrm{d}r^2 +r^2\hat{g}_{ij}\mathrm{d}x^i\mathrm{d}x^j\,,   \qquad   \text{$\hat{g}_{ij}$ being the metric of Sol, Nil or $SL_2R$ manifold,}     \\
&&   U(\phi) = \sigma \phi^4 \,,   \phi(r) =  \dfrac{A}{r^{3/2}}   \,,  V(r)=\dfrac{r^2}{4\alpha} - \mu - \dfrac{3\kappa A^2}{64\alpha r} \,, \qquad  \alpha = -\frac{3}{4\Lambda}\,,    \label{eq:nu} \sigma = -\dfrac{27\kappa}{2048\alpha}\,,  \xi=1 \nonumber \,.
\end{eqnarray}
\textbf{In this solution the horizon gemetry can be the 3 types of Thurston geometries, Sol, Nil or $SL_2R$.}
This solution is just (\ref{BHvacuum}) with the matter field added, and the horizon geometry is somehow restricted, compared to the totally arbitrary horizon in (\ref{BHvacuum}).
One might feel really curious about how the horizon geometry gets restricted after adding matter.
This section will explain this.
\newline\newline
\textbf{To be specific, we are going to show that, for the model (\ref{GBactionMat})
\newline
1. How the arbitrariness of $\hat{g}_{ij}$ in (\ref{BHvacuum}) comes out.
\newline
\newline
2. In (\ref{BHScalar}), the horizon metric $\hat{g}_{ij}$ is not restricted to the 3 types of Thurston geometries, but can be arbitrary 3-dimensional metric with CONSTANT scalar curvature $\hat{R}$.
\newline
\newline
3. To add an $r$-dependent scalar field $\phi(r)$ to the background (\ref{BHvacuum}), the only possibility is the solution (\ref{BHScalar}) with an arbitrary $\hat{g}_{ij}$ possessing constant $\hat{R}$.}
\newline\newline
\textbf{Here comes the proof.}
For $n=3$, in the presence of $\phi(r)$ the gravitational field equations (\ref{GBeomphi}) are shown below,
\begin{eqnarray}
\textbf{$tt$ eq.}&&\nonumber\\
&&\hat{R} \left(-3 \kappa  \xi  r \phi (r)^2-32 \alpha  {V}'(r)+16 r\right)\nonumber\\
&&-32 \Lambda  r^3-32 \kappa  r^3 {U}(\phi (r))+6 \kappa  \xi  r^3 \phi (r) {V}'(r) \phi '(r)+9 \kappa  \xi  r^2 \phi (r)^2 {V}'(r)-48 r^2 {V}'(r)\nonumber\\
&&\qquad+2 {V}(r) \Big(6 \kappa  \xi  r^2 \phi (r) \left(3 \phi '(r)+r \phi ''(r)\right)+6 \kappa  \xi  r^3 \phi '(r)^2-8 \kappa  r^3 \phi '(r)^2+9 \kappa  \xi  r \phi (r)^2+96 \alpha  {V}'(r)-48 r \Big)=0\,,\nonumber\\
\label{tteq}\\
\textbf{$rr$ eq.}&&\nonumber\\
&&\hat{R} \left(-3 \kappa  \xi  r \phi (r)^2-32 \alpha  {V}'(r)+16 r\right)\nonumber\\
&&-32 \Lambda  r^3-32 \kappa  r^3 {U}(\phi (r))+6 \kappa  \xi  r^3 \phi (r) {V}'(r) \phi '(r)+9 \kappa  \xi  r^2 \phi (r)^2 {V}'(r)-48 r^2 {V}'(r)
\nonumber\\
&&\qquad+2 {V}(r) \Big(18 \kappa  \xi  r^2 \phi (r) \phi '(r) +8 \kappa  r^3 \phi '(r)^2+9 \kappa  \xi  r \phi (r)^2+96 \alpha  {V}'(r)-48 r \Big)=0\,,\label{rreq}\\
\textbf{$ij$ eq.}&&\nonumber\\
&&(\hat{R}_{ij}-\frac{1}{2}\hat{R} \hat{g}_{ij}) \left(-3 \kappa  \xi  \phi (r)^2-32 \alpha  {V}''(r)+ 16\right)\nonumber\\
&&+ \hat{g}_{ij}\Big[16 \Lambda  r^2+16 \kappa  r^2 {U}(\phi (r))
-6 \kappa  \xi  r^2 \phi (r) {V}'(r) \phi '(r)-\frac{3}{2} \kappa  \xi  r^2 \phi (r)^2 {V}''(r)+8 r^2 {V}''(r)-32 \alpha  {V}'(r)^2-6 \kappa  \xi  r \phi (r)^2 {V}'(r)\nonumber\\
&&+32 r {V}'(r)-{V}(r) \left(2 \kappa  (3 \xi -4) r^2 \phi '(r)^2+6 \kappa  \xi  r \phi (r) \left(2 \phi '(r)+r \phi ''(r)\right)+3 \kappa  \xi  \phi (r)^2+32 \alpha  {V}''(r)-16\right)\Big]=0\label{ijeq}\,,
\end{eqnarray}
and the scalar field equation is
\begin{eqnarray}
    -U'(\phi)+\frac{3 \xi  \phi (r) \left(-\hat{R}+r^2 {V}''(r)+6 r {V}'(r)+6 {V}(r)\right)}{16 r^2}+{V}'(r) \phi '(r)+{V}(r) \left(\frac{3 \phi '(r)}{r}+\phi ''(r)\right)=0\,.
    \label{scalareq}
\end{eqnarray}
\newline\newline
\textbf{Proof 1. The arbitrariness of $\hat{g}_{ij}$ for the vaccuum case.}

In the equations (\ref{tteq}), (\ref{rreq}) and (\ref{ijeq}), if $\phi=0$, the only terms related to the 3-dimensional geometry are 
\begin{eqnarray}
    \hat{R} \left(-32 \alpha  {V}'(r)+16 r\right) \quad \text{and} \quad (\hat{R}_{ij}-\frac{1}{2}\hat{R} \hat{g}_{ij}) \left(-32 \alpha  {V}''(r)+ 16\right)\,.
\end{eqnarray}
Obviously when we choose
\begin{eqnarray}\label{Vforallvanish}
    V(r) = \frac{r^2}{4\alpha}-\mu \qquad \text{ where $\mu$ is an integration constant and positive for a black hole,}
\end{eqnarray}
\textbf{the $\hat{R}, \hat{R}_{ij}$ and $\hat{g}_{ij}$ terms all dissapear, and then the 3-dimensional geometry does not appear in any of these equations.}
Under this condition the $tt$, $rr$ and $ij$ equations become
\begin{eqnarray}
    -(4 \alpha  \Lambda +3) \left(r^2-4 \alpha  \mu \right)/(16 \alpha ^2)\,, \qquad \hat{g}_{ij} r^2 (4 \alpha  \Lambda +3)/(4 \alpha )\,.
\end{eqnarray}
The choice $\alpha = - 3/(4\Lambda) $ will satisfy these equations. 
This is why the vacuum model admits an arbitrary $\hat{g}_{ij}$.
\newline\newline
\textbf{Proof 2. The arbitrariness of $\hat{g}_{ij}$ in the presence of $\phi(r)$.}
When $\phi(r) \neq 0$ the terms related to $\hat{g}_{ij}$ are
\begin{eqnarray}
    \hat{R} \left(-3 \kappa  \xi  r \phi (r)^2-32 \alpha  {V}'(r)+16 r\right) \quad \text{and} \quad (\hat{R}_{ij}-\frac{1}{2}\hat{R} \hat{g}_{ij}) \left(-3 \kappa  \xi  r \phi (r)^2-32 \alpha  {V}''(r)+ 16\right)\,.
\end{eqnarray}

Different from the vacuum case above with the choice (\ref{Vforallvanish}), these terms can not vanish simultaneously for any $V(r)$ and nonzero $\phi(r)$.
\textbf{The solution (\ref{BHScalar}) above ensures that the term with $(\hat{R}_{ij}-\frac{1}{2}\hat{R} \hat{g}_{ij})$ in the $ij$  equations dissapears, but the terms with $\hat{R}$ in the $tt$ and $rr$ equations  still exist and the separation of variables in these equations forces $\hat{R}$ to be a constant.
This is the only constraint on the base manifold geometry in their model.}
So $\hat{g}_{ij}$ can be beyond the metric of Sol, Nil or $SL_2R$ manifold, but ANY manifold with constant scalar curvature $\hat{R}$.
\newline\newline
\textbf{Proof 3. The uniqueness of the solution (\ref{BHScalar}). }

To ``add'' scalar field into the vacuum background (\ref{BHvacuum}) we must treat the scalar field $\phi(r)$ as a small perturbation first and then give the nonperturbative analysis.
The lapse function $V(r)$ is set to include a linear perturbation $\beta(r)$:
\begin{eqnarray}
V(r)=\dfrac{r^2}{4\alpha} - \mu + \beta(r) \,, \qquad  \alpha = -\frac{3}{4\Lambda}\,.
\end{eqnarray}
To maintain some arbitrariness of $\hat{g}_{ij}$, the coefficient of $(\hat{R}_{ij}-\frac{1}{2}\hat{R} \hat{g}_{ij}) \left(-3 \kappa  \xi  r \phi (r)^2-32 \alpha  {V}''(r)+ 16\right)$ should vanish\footnote{Otherwise this geometry has to be Einsteinian satsfying $\hat{R}_{ij} \propto \hat{g}_{ij}$ due to separation of variables.}, leading to
\begin{eqnarray}
\phi(r)=2 \sqrt{\frac{2}{3}}\frac{ \sqrt{2 \Lambda +3 \mathcal{V}''(r)}}{\sqrt{\kappa } \sqrt{\Lambda } \sqrt{\xi }} = \frac{2 \sqrt{2} \sqrt{\beta ''(r)}}{\sqrt{\kappa } \sqrt{\Lambda } \sqrt{\xi }}.
\end{eqnarray}
The perturbative form of $U(\phi)$ can be written as
\begin{eqnarray}
    U(\phi) = c_0 + c_1 \phi + c_2 \phi^2 + c_3 \phi^3 + c_4\phi^4 + \text{higher order terms}.
\end{eqnarray}
In the gravitational field equations, we need to treat those terms quadratic in $\phi(r)$ as first-order perturbations --- that is to say, terms linear in $\beta(r)$.
Keeping the terms up to terms linear in $\beta$, the combined equation 
 \quad $(\ref{tteq})/V(r) + (\ref{ijeq})/(r^2\hat{g}_{ij})$ \quad forces
\begin{eqnarray}
    \beta(r) = \epsilon\, r^{-\frac{\hat{R}}{2 \mu }-2}
    \label{beta}
\end{eqnarray}
with $\epsilon$ denoting the linear perturbation.
Putting (\ref{beta}) into (\ref{tteq}) linear in $\beta$ gives
\begin{eqnarray}
\hat{R} = \frac{2 \mu  (8-9 \xi )}{3 \xi -2} \,, \qquad  c_2= -\frac{3 \left(\Lambda  \xi ^2-\Lambda  \xi \right)}{4 (3 \xi -2)}\,, \qquad c_0=c_1=0.
\end{eqnarray}
The equation (\ref{scalareq}) is expanded into terms linear and cubic in $\sqrt{\beta}$.
Its linear part forces $\xi=1$, then we have $\hat{R}=-2\mu$, $\beta(r) = \epsilon r^{-1}$, and $c_2=0$.
The higher order terms of the equations (\ref{tteq}), (\ref{rreq}), (\ref{ijeq}) and (\ref{scalareq}) become simple and will force $c_3=0$ and $512 {c_4}-9 \kappa  \Lambda=0$. That is to say, $U(\phi) = c_4 \phi^4$, $c_4=-{27\kappa}/({2048\alpha})$.
\textbf{All these combined will give the solution (\ref{BHScalar}) with $\hat{R}=-2\mu$. This is the uniqueness of this solution.}
And one can verify easily that when $\epsilon$ is not small, the full equations of motion are satisfied too.

\section{$F(R)$ black hole}\label{sectionofFR}
We now introduce a vacuum case similar to (\ref{BHvacuum}) in generic $F(R)$ theory with the action
\begin{eqnarray}
S=\frac{1}{2\kappa }\int \mathrm{d}^{D} x \sqrt{-g} F(R)
\end{eqnarray}
where the Lagrangian is a function of the Ricci scalar $R$.
The equations of motion are
\begin{eqnarray}\label{eomFR}
    \left(R_{\mu\nu}+ g_{\mu\nu}\Box-\nabla_\mu \nabla_\nu \right)F'(R) - \frac{1}{2}F(R) g_{\mu\nu}=0\,,
\end{eqnarray}
with $F'(R)=\mathrm{d}F(R)/\mathrm{d}R$.

This theory admit a simple class of solutions when the spacetime Ricci scalar $R$ is a constant and $F(R)=F'(R)=0$.
In this case the field equations (\ref{eomFR}) are automatically satisfied.
For the ansatz $(\ref{decomp})$, according to the decompostion of the curvature tensor in the appendix, the Ricci scalar $\hat{R}$ must be a constant since $R$ is a constant. The only term including the Ricci tensor $\hat{R}_{ij}$ vanishes since $F'(R)=0$. \textbf{Therefore $\hat{g}_{ij}$ is an arbitrary metric with constant $\hat{R}$, quite similar to the vacuum case (\ref{BHvacuum}) in the Einstein-Gauss-Bonnet theory.}

For this solution, the Wald entropy is automatically $0$:
\begin{eqnarray}
    S\equiv\frac{A}{4}F'(R)|_\text{horizon}=0\,.
\end{eqnarray}
since $F'(R)=0$.
The vanishing entropy at some critical coupling is not a new phenomenon and it has been discovered in $R^2$ gravity in $4$ and $5$ dimensions \cite{Cai:2009ac,Liu:2012yd} for Lifshitz black holes.
Similar effect was reported in new massive gravity \cite{Liu:2009bk,Liu:2009kc}.

At the end we provide an simple example of quadratic curvature gravity, $F(R)=R+\lambda R^2 - 2\Lambda$.
The condition $F'(R)=0$ leads to
\begin{eqnarray}\label{BHSolution}
V(r)=c+\frac{r^2}{2 \lambda (n+1)(n+2)} - \frac{\mu _1}{r^{n-1}} - \frac{\mu _2}{r^n}
\end{eqnarray}
with $\mu_1$ and $\mu_2$ constants of integration, and $c$ proportional to the constant $\hat{R}$.
The condition $F(R) = 0$ forces the coupling constant
\begin{eqnarray}\label{criticalcoupling}
    \lambda = -\frac{1}{8\Lambda}\,.
\end{eqnarray}
This critical point was pointed out in Ref.\cite{Nojiri:2001aj} at which the deconfinement-confinement phase structure is reversed.
This critical coupling constant is very much similar to the $\alpha=-3/(4\Lambda)$ in the Einstein-Gauss-Bonnet case.

Hervik and Ortaggio\cite{Hervik:2019gly} mentioned a solution with the function $V(r)$ as in (\ref{BHSolution}), and they were considering the ``universal'' horizon case.
The horizon is called universal if any symmetric conserved rank-$2$ tensor constructed from sums of terms involving contractions of the metric and powers of arbitrary covariant derivatives of the curvature tensor is proportional to the metric $\hat{g}_{ij}$.
\textbf{We first point out that the horizon only needs to have constant Ricci scalar, not to be universal.}
An important example is just the Sol manifold, where it is not universal but has constant Ricci scalar.

\section{conclusion}
In this paper we explore the effect of matter fields on black hole horizon geometry.

In Sec.\ref{GBsection} we reexamined 2 classes of static black hole solutions with nontrivial horizon geometry (other than maximally symmetric or Einstein) in 5-dimensional Einstein-Gauss-Bonnet
gravity theory with a critical coupling constant $\alpha=-3/(4\Lambda)$, i.e. the Chern-Simons gravity theory.
The first vacuum solution, first proposed by \cite{Dotti:2007az}, admits a horizon geometry that is arbitrary, and we briefly showed how this arbitrariness come out from the equations of motion.
This kind of explanation can already be found in the paper \cite{Dotti:2010bw} where the formulation of vielbeins was applied.
Our explanation used the traditional coordinate system language, and it is easy to extend this kind of analysis to the cases including matter fields.
Applying this kind of analysis, we explore the effect of matter fields on the horizon geometry --- the second class of solutions are black holes with scalar hair provided by the recent paper \cite{Guajardo:2024hrl}. 
These solutions are just the the same first solution except for the scalar matter added.
The horizon geometry gains some new restriction.
In \cite{Guajardo:2024hrl} the black hole horizons appear in the form of 3 Thurston geometries: Sol, Nil and $SL_2R$.
In this paper we proved that the horizon manifold can be beyond these 3 types --- the only restriction from the scalar field is $\hat{R}$ being a constant. The horizon still maintains much arbitrariness.
We also proved a uniqueness theorem of this model.

In Sec.\ref{sectionofFR} we proved that such arbitrary horizon geometry with constant Ricci scalar appears in generic $F(R)$ gravity theory.
Further more, we showed a specific example with a quadratic correction, $F(R) = R+\lambda R^2 - 2\Lambda$.
Similar to the Einstein-Gauss-Bonnet case, this specific solution has a critical coupling $\lambda=-1/(8\Lambda)$.

Now we make the final summary. Given the evidence in Einstein-Gauss-Bonnet theory and $F(R)$ theory, a critical coupling constant may be a common feature for the black holes with nontrivial unconstrained horizon geometry in higher order gravity theories.
Moreover, the matter fields added into the black hole background may provide some restriction to the unconstrained horizon.
This is shown in the Einstein-Gauss-Bonnet case, while in the $F(R)$ case we hope to find similar evidence in the future.

%One can easily verify that the bulk part of the on-shell action (\ref{Action}) of this solution is constantly $0$ since the Ricci scalar $R=4 \Lambda$ and $\lambda=-1/(8\Lambda)$:
%\begin{eqnarray}
%I_{\text{on-shell}}=0\,.
%\end{eqnarray}

\section{Acknowledgements}
I especially thank Prof. Rong-Gen Cai for supporting me exploring this subject.
And I would like to thank Prof. Julio Oliva for extremely helpful discussions.
Besides, I would like to thank Dr. Zu-Cheng Chen for his hospitality during my visit at Hunan Normal University, and to thank Prof. De-Cheng Zou, Prof. Yi-Bin Huang and Dr. Tong Chern for helpful discussions.
This work was supported by the National Natural Science Foundation of China with Grant No.12265001 and the East China University of Technology Research Foundation for Advanced Talents (No. DHBK2019198).

\section{Appendix}

We now show the decomposition of the curvature tensors of the warped product (\ref{decomp}), and then provide the expressions for the field equations in the Einstein-Gauss-Bonnet theory.
The decompostion of the curvature tensors are
\begin{eqnarray}
    &&{R_{abc}}^d={\bar{R}\vphantom{R}_{abc}}^d\,,\qquad {R_{aib}}^j = -\frac{\bar{\nabla}_a\bar{\nabla}_b\gamma}{\gamma}\delta_j^i\,,\\
    &&{R_{ijk}}^l={\hat{R}\vphantom{R}_{ijk}}^l - \bar{\nabla}_c\gamma \bar{\nabla}^c\gamma( \delta_j^l\hat{g}_{ki} - \delta_i^l\hat{g}_{kj})\,,\\
    &&R_{ab}=\bar{R}_{ab} - n \frac{\bar{\nabla}_a\bar{\nabla}_b\gamma}{\gamma}\,,\\
    &&R_{ij}=\hat{R}_{ij}+X(u)\hat{g}_{ij}\,,\qquad X = -\gamma \bar{\nabla} _c\bar{\nabla} ^c\gamma -(n-1)\bar{\nabla} _c\gamma \bar{\nabla} ^c\gamma\,,\\
    &&R=\bar{R}-2n\frac{\bar{\nabla}_c\bar{\nabla}^c\gamma}{\gamma}-n(n-1)\frac{\bar{\nabla} _c\bar{\nabla} ^c\gamma}{\gamma^2}+\frac{1}{\gamma^2}\hat{R}\,.
\end{eqnarray}

The action of $D$-dimensional EGB gravity theory with a negative cosmological constant
\begin{eqnarray}\label{GBactionVac}
    S &= &\frac{1}{2\kappa }\int \mathrm{d}^{D} x \sqrt{-g}\left[ R - 2\Lambda + \alpha (R^2 -4 R_{\mu\nu}R^{\mu\nu} + R_{\mu\nu \rho \sigma}R^{\mu\nu \rho \sigma}) \right]
    %\nonumber\\
    %&&-\frac{1}{4}\int \mathrm{d}^{d+1} x \sqrt{-g}F_{\mu\nu}F^{\mu\nu}
    \,,
\end{eqnarray}
where the cosmological constant $\Lambda<0$, and the Gauss-Bonnet coupling denoted by $\alpha$. The equations of motion are
\begin{eqnarray}
    &&R_{\mu\nu} - \frac{1}{2}R g_{\mu\nu} + \Lambda g_{\mu\nu}\nonumber\\
    &&- \alpha\left(4 R_{\mu\rho} R^\rho\,_\nu - 2R R_{\mu\nu}+ 4 R^{\rho \sigma}R_{\mu\rho\nu\sigma} - 2 R_\mu\,^{\rho\sigma\tau}R_{\nu\rho\sigma\tau} + \frac{1}{2} g_{\mu\nu} \left(R^2 -4 R_{\rho \sigma}R^{\rho \sigma} + R_{\rho\sigma\tau\pi}R^{\rho\sigma\tau\pi}\right)\right)\nonumber\\
    &&=0
    %&&\frac{1}{2}(F_{\mu\rho}F_\nu{}^{\rho}-\frac{1}{4}F_{\rho \sigma}F^{\rho \sigma}g_{\mu\nu})
    \,.\label{GBeom}
\end{eqnarray}

Firstly we look at the equations (\ref{GBeom}) decomposed into $ab$ and $ij$ components:
\begin{eqnarray}
    &&R_{a\rho} R^\rho\,_b=R_{ac} R^c\,_b = \bar{R}_{ac} \bar{R}^c\,_b - n \bar{R}_{ac} \frac{\bar{\nabla}^c\bar{\nabla}_b\gamma}{\gamma} - n \bar{R}_{bc} \frac{\bar{\nabla}_a\bar{\nabla}^c\gamma}{\gamma} + n^2 \frac{\bar{\nabla}_a\bar{\nabla}_c\gamma \bar{\nabla}^c\bar{\nabla}_b\gamma}{\gamma^2} \,,\\
    &&R R_{ab} = (\bar{R}-2n\frac{\bar{\nabla}_c\bar{\nabla}^c\gamma}{\gamma}-n(n-1)\frac{\bar{\nabla} _c\bar{\nabla} ^c\gamma}{\gamma^2}+\frac{1}{\gamma^2}\hat{R})(\bar{R}_{ab} - n \frac{\bar{\nabla}_a\bar{\nabla}_b\gamma}{\gamma})\,,\\
    &&R^{\rho \sigma}R_{a\rho b\sigma} = R^{c d}\bar{R}_{a c b d} + R^{i j}{R}_{a i b j}
    = R^{c d}\bar{R}_{a c b d} -\gamma^{-3} \bar{\nabla}_a\bar{\nabla}_b\gamma (\hat{R}+nX(u) )\,,\\
    &&R_a\,^{\rho\sigma\tau}R_{b\rho\sigma\tau} = \bar{R}_a\,^{cde}\bar{R}_{bcde} +2 R_a\,^{icj}R_{bicj}
    =  \bar{R}_a\,^{cde}\bar{R}_{bcde} +2 n \frac{1}{\gamma^2} \bar{\nabla}_a\bar{\nabla}_c\gamma \bar{\nabla}^c\bar{\nabla}_b\gamma\,,\\
    &&R_{i\rho} R^\rho\,_j=R_{ik} R^k\,_j = \gamma^{-2}[\hat{R}_{ik} \hat{R}^k\,_j + 2  X(u) \hat{R}_{ij} +X(u)^2 \hat{g}_{ij}]
    \,,\\
    &&R R_{ij} = (\bar{R}-2n\frac{\bar{\nabla}_c\bar{\nabla}^c\gamma}{\gamma}-n(n-1)\frac{\bar{\nabla} _c\bar{\nabla} ^c\gamma}{\gamma^2}+\frac{1}{\gamma^2}\hat{R})(\hat{R}_{ij}+X(u)\hat{g}_{ij})\,,\\
    &&R^{\rho \sigma}R_{i\rho j\sigma} = R^{c d}{R}_{i c j d} + R^{k l}{R}_{i k j l}\nonumber\\
    &&=-\gamma R^{cd} \bar{\nabla}_c\bar{\nabla}_d\gamma \hat{g}_{ij} + \gamma^{-2}\big[    \hat{R}^{k l}\hat{R}_{ik jl} -  (n-1) \bar{\nabla}_c\gamma \bar{\nabla}^c\gamma X(u) \hat{g}_{ij}  + X(u)\hat{R}_{ij} - \bar{\nabla}_c \gamma \bar{\nabla}^c\gamma (\hat{R}\hat{g}_{ij} - \hat{R}_{ij})  \big]       \,,\\
    &&R_i\,^{\rho\sigma\tau}R_{j\rho\sigma\tau} = 2 \bar{\nabla}_c\bar{\nabla}_d\gamma \bar{\nabla}^c\bar{\nabla}^d\gamma \hat{g}_{ij}  +  \gamma^{-2} \Big[  \hat{R}_i\,^{klm}\hat{R}_{jklm} - 4   \bar{\nabla}_c\gamma \bar{\nabla}^c \gamma \hat{R}_{ij}  + 2 (n-1) (\bar{\nabla}_c\gamma \bar{\nabla}^c \gamma)^2 \hat{g}_{ij}        \Big] \,,    \\
    &&R^2 = (\bar{R}-2n\frac{\bar{\nabla}_c\bar{\nabla}^c\gamma}{\gamma}-n(n-1)\frac{\bar{\nabla} _c\bar{\nabla} ^c\gamma}{\gamma^2}+\frac{1}{\gamma^2}\hat{R})^2\,,\\
    && R_{\rho \sigma}R^{\rho \sigma} = R^{cd} R_{cd} + R^{ij}R_{ij} = R^{cd} R_{cd} + \gamma^{-4}(\hat{R}^{ij}\hat{R}_{ij} + 2\hat{R}X(u)+n X(u)^2)\,,\\
    && R_{\rho\sigma\tau\pi}R^{\rho\sigma\tau\pi} = \bar{R}_{abcd}\bar{R}^{abcd} + 4R_{aibj}R^{aibj}+R_{ijkl}R^{ijkl}\\
    && = \bar{R}_{abcd}\bar{R}^{abcd} + \frac{4n}{\gamma^2} \bar{\nabla}_a\bar{\nabla}_b\gamma \bar{\nabla}^a\bar{\nabla}^b\gamma + \gamma^{-4} \big[\hat{R}_{ijkl}\hat{R}^{ijkl} -4 \bar{\nabla}_c\gamma \bar{\nabla}^c\gamma \hat{R} + 2n(n-1)(\bar{\nabla}_c\gamma \bar{\nabla}^c\gamma)^2\big]\,.
\end{eqnarray}


\begin{thebibliography}{99}
	

%\cite{Birmingham:1998nr}
\bibitem{Birmingham:1998nr}
D.~Birmingham,
%``Topological black holes in Anti-de Sitter space,''
Class. Quant. Grav. \textbf{16}, 1197-1205 (1999)
doi:10.1088/0264-9381/16/4/009
[arXiv:hep-th/9808032 [hep-th]].
%399 citations counted in INSPIRE as of 13 Dec 2020

\bibitem{GibWil87}
G.~W. Gibbons and D.~L. Wiltshire, {\it Space-time as a membrane in higher
    dimensions},  {\em Nucl. Phys. {\rm B}} {\bf 287} (1987) 717--742.


%\cite{Yang:2023nnk}
\bibitem{Yang:2023nnk}
J.~Yang,
%``Novel topological black holes from thermodynamics and deforming horizons,''
[arXiv:2301.01709 [gr-qc]].
%0 citations counted in INSPIRE as of 16 Dec 2023


\bibitem{Ryan:1975jw} 
M.~P.~Ryan and L.~C.~Shepley,
%``Homogeneous Relativistic Cosmologies,''
Princeton, Usa: Univ. Pr. ( 1975) 320 P. ( Princeton Series In Physics)


\bibitem{Th1} W.P.\ Thurston, {\it Three-Dimensional Geometry and 
    Topology}, ed.\ S.\ Levy (Princeton University Press, Princeton, 1997).

\bibitem{Sc1} P.\ Scott, {\it Bull.\ London Math.\ Soc.}\ {\bf 15} (1983),
401. 














%\cite{Cadeau:2000tj}
\bibitem{Cadeau:2000tj}
C.~Cadeau and E.~Woolgar,
%``New five-dimensional black holes classified by horizon geometry, and a Bianchi VI brane world,''
Class. Quant. Grav. \textbf{18}, 527-542 (2001).
%34 citations counted in INSPIRE as of 04 Oct 2020		



%\cite{Hervik:2003vx}
\bibitem{Hervik:2003vx}
S.~Hervik,
%``Einstein metrics: Homogeneous solvmanifolds, generalized Heisenberg groups and black holes,''
J. Geom. Phys. \textbf{52}, 298-312 (2004)
doi:10.1016/j.geomphys.2004.03.005
[arXiv:hep-th/0311108 [hep-th]].
%13 citations counted in INSPIRE as of 22 Mar 2024



%\cite{Arias:2017yqj}
\bibitem{Arias:2017yqj}
R.~E.~Arias and I.~Salazar Landea,
%``Thermoelectric Transport Coefficients from Charged Solv and Nil Black Holes,''
JHEP \textbf{12}, 087 (2017)
doi:10.1007/JHEP12(2017)087
[arXiv:1708.04335 [hep-th]].
%13 citations counted in INSPIRE as of 22 Mar 2024

%\cite{Arias:2018mqn}
\bibitem{Arias:2018mqn}
R.~E.~Arias and I.~Salazar Landea,
%``Intermediate scalings for Solv, Nil and $SL_2({\cal R})$ black branes,''
Phys. Rev. D \textbf{99}, no.10, 106015 (2019)
doi:10.1103/PhysRevD.99.106015
[arXiv:1812.09108 [hep-th]].
%4 citations counted in INSPIRE as of 22 Mar 2024

%\cite{Faedo:2019rgo}
\bibitem{Faedo:2019rgo}
F.~Faedo, D.~A.~Farotti and S.~Klemm,
%``Black holes in Sol minore,''
JHEP \textbf{12}, 151 (2019)
doi:10.1007/JHEP12(2019)151
[arXiv:1908.07421 [hep-th]].
%2 citations counted in INSPIRE as of 22 Mar 2024


%\cite{Figueroa:2021apr}
\bibitem{Figueroa:2021apr}
J.~Figueroa and M.~Oyarzo,
%``Slowly rotating black holes modeled by Solv geometries,''
[arXiv:2102.02328 [gr-qc]].
%0 citations counted in INSPIRE as of 22 Mar 2024



%\cite{Peng:2021xwh}
\bibitem{Peng:2021xwh}
Y.~Peng,
%``New topological Gauss-Bonnet black holes in five dimensions,''
Phys. Rev. D \textbf{104}, no.8, 084004 (2021)
doi:10.1103/PhysRevD.104.084004
[arXiv:2105.08482 [gr-qc]].
%3 citations counted in INSPIRE as of 20 Mar 2024



%\cite{Naderi:2021skd}
\bibitem{Naderi:2021skd}
F.~Naderi, A.~Rezaei-Aghdam and Z.~Mahvelati-Shamsabadi,
%``Spatially homogeneous black hole solutions in $z=4$ Ho\v{r}ava\textendash{}Lifshitz gravity in $(4+1)$ dimensions with Nil geometry and $H^2\times R$ horizons,''
Eur. Phys. J. C \textbf{81}, no.10, 865 (2021)
doi:10.1140/epjc/s10052-021-09622-7
[arXiv:2106.03217 [hep-th]].
%1 citations counted in INSPIRE as of 22 Mar 2024




%\cite{Faedo:2022hle}
\bibitem{Faedo:2022hle}
F.~Faedo, S.~Klemm and P.~Mariotti,
%``Rotating black holes with Nil or SL(2, \ensuremath{\mathbb{R}}) horizons,''
JHEP \textbf{05}, 138 (2023)
doi:10.1007/JHEP05(2023)138
[arXiv:2212.04890 [hep-th]].
%1 citations counted in INSPIRE as of 22 Mar 2024





%\cite{Dotti:2007az}
\bibitem{Dotti:2007az}
G.~Dotti, J.~Oliva and R.~Troncoso,
%``Exact solutions for the Einstein-Gauss-Bonnet theory in five dimensions: Black holes, wormholes and spacetime horns,''
Phys. Rev. D \textbf{76}, 064038 (2007)
doi:10.1103/PhysRevD.76.064038
[arXiv:0706.1830 [hep-th]].
%102 citations counted in INSPIRE as of 03 Jun 2021


%\cite{Dotti:2010bw}
\bibitem{Dotti:2010bw}
G.~Dotti, J.~Oliva and R.~Troncoso,
%``Static solutions with nontrivial boundaries for the Einstein-Gauss-Bonnet theory in vacuum,''
Phys. Rev. D \textbf{82}, 024002 (2010)
doi:10.1103/PhysRevD.82.024002
[arXiv:1004.5287 [hep-th]].
%29 citations counted in INSPIRE as of 05 Apr 2024


%\cite{Oliva:2012ff}
\bibitem{Oliva:2012ff}
J.~Oliva,
%``All the solutions of the form M2(warped)x\textbackslash{}Sigma(d-2) for Lovelock gravity in vacuum in the Chern-Simons case,''
J. Math. Phys. \textbf{54}, 042501 (2013)
doi:10.1063/1.4795258
[arXiv:1210.4123 [hep-th]].
%15 citations counted in INSPIRE as of 05 Apr 2024


%\cite{Guajardo:2024hrl}
\bibitem{Guajardo:2024hrl}
L.~Guajardo and J.~Oliva,
%``Primary scalar hair in Gauss\textendash{}Bonnet black holes with Thurston horizons,''
Eur. Phys. J. C \textbf{85}, no.2, 139 (2025)
doi:10.1140/epjc/s10052-025-13869-9
[arXiv:2412.20134 [hep-th]].
%1 citations counted in INSPIRE as of 28 Feb 2025









%\cite{Chamseddine:1989nu}
\bibitem{Chamseddine:1989nu}
A.~H.~Chamseddine,
%``Topological Gauge Theory of Gravity in Five-dimensions and All Odd Dimensions,''
Phys. Lett. B \textbf{233}, 291-294 (1989)
doi:10.1016/0370-2693(89)91312-9
%231 citations counted in INSPIRE as of 22 Mar 2024


%\cite{Brihaye:2013vsa}
\bibitem{Brihaye:2013vsa}
Y.~Brihaye and E.~Radu,
%``Black hole solutions in d=5 Chern-Simons gravity,''
JHEP \textbf{11}, 049 (2013)
doi:10.1007/JHEP11(2013)049
[arXiv:1305.3531 [gr-qc]].
%12 citations counted in INSPIRE as of 22 Mar 2024





%\cite{Cai:2009ac}
\bibitem{Cai:2009ac}
R.~G.~Cai, Y.~Liu and Y.~W.~Sun,
%``A Lifshitz Black Hole in Four Dimensional R**2 Gravity,''
JHEP \textbf{10}, 080 (2009)
doi:10.1088/1126-6708/2009/10/080
[arXiv:0909.2807 [hep-th]].
%124 citations counted in INSPIRE as of 09 Apr 2024



%\cite{Liu:2012yd}
\bibitem{Liu:2012yd}
Y.~Liu,
%``Spatially homogeneous Lifshitz black holes in five dimensional higher derivative gravity,''
JHEP \textbf{06}, 024 (2012)
doi:10.1007/JHEP06(2012)024
[arXiv:1202.1748 [hep-th]].
%9 citations counted in INSPIRE as of 09 Apr 2024



%\cite{Liu:2009bk}
\bibitem{Liu:2009bk}
Y.~Liu and Y.~w.~Sun,
%``Note on New Massive Gravity in AdS(3),''
JHEP \textbf{04}, 106 (2009)
doi:10.1088/1126-6708/2009/04/106
[arXiv:0903.0536 [hep-th]].
%97 citations counted in INSPIRE as of 09 Apr 2024


%\cite{Liu:2009kc}
\bibitem{Liu:2009kc}
Y.~Liu and Y.~W.~Sun,
%``Consistent Boundary Conditions for New Massive Gravity in $AdS_3$,''
JHEP \textbf{05}, 039 (2009)
doi:10.1088/1126-6708/2009/05/039
[arXiv:0903.2933 [hep-th]].
%83 citations counted in INSPIRE as of 09 Apr 2024


%\cite{Nojiri:2001aj}
\bibitem{Nojiri:2001aj}
S.~Nojiri and S.~D.~Odintsov,
%``Anti-de Sitter black hole thermodynamics in higher derivative gravity and new confining deconfining phases in dual CFT,''
Phys. Lett. B \textbf{521}, 87-95 (2001)
[erratum: Phys. Lett. B \textbf{542}, 301 (2002)]
doi:10.1016/S0370-2693(01)01186-8
[arXiv:hep-th/0109122 [hep-th]].
%166 citations counted in INSPIRE as of 20 Mar 2024



%\cite{Hervik:2019gly}
\bibitem{Hervik:2019gly}
S.~Hervik and M.~Ortaggio,
%``Universal Black Holes,''
JHEP \textbf{02}, 047 (2020)
doi:10.1007/JHEP02(2020)047
[arXiv:1907.08788 [gr-qc]].
%11 citations counted in INSPIRE as of 20 Mar 2024

		
	\end{thebibliography}
\end{document}